\documentclass[11pt]{article}

\usepackage{acl} 

\usepackage[T1]{fontenc}
\usepackage{graphicx}
\usepackage{enumerate}
\usepackage{xcolor}

\usepackage{epsfig}
\usepackage{graphicx}
\usepackage{amssymb}
\usepackage{amsmath}
\usepackage[utf8]{inputenc}
\usepackage[T1]{fontenc}
\usepackage[english]{babel}
\usepackage{fancyhdr}
\usepackage{mdframed}
\usepackage{newtxtext,newtxmath} 
\usepackage{hyperref}
\usepackage{textcomp}
\usepackage{tablefootnote}
\usepackage[subtle]{savetrees}
\usepackage[hang,flushmargin]{footmisc}

\usepackage{microtype}
\usepackage{subcaption}
\usepackage{graphicx}
\usepackage{csquotes}
\usepackage{amssymb}

\usepackage{amsthm}
\usepackage{pifont}
\usepackage{verbatim}
\usepackage{listings}
\usepackage{booktabs} 

\usepackage{algorithm}
\usepackage{algpseudocode}

\usepackage{hyperref}
\usepackage{enumitem}
\usepackage{tabularx}
\usepackage{float}
\usepackage{multirow}
\usepackage{listings}
\usepackage{xcolor}
\lstdefinestyle{prompt}{
  basicstyle=\ttfamily\small,
  backgroundcolor=\color{gray!10},
  frame=single,
  breaklines=true
}
\usepackage[most]{tcolorbox}  

\newtheorem{definition}{Definition}
\newtheorem{problem}{Problem Statement}

\usepackage{graphicx}

\title{Towards \textit{Contextual} Sensitive Data Detection}

\author{\textbf{Liang Telkamp}\textsuperscript{1,2} and \textbf{Madelon Hulsebos\textsuperscript{2}\thanks{Correspondence to \href{mailto:madelon@cwi.nl}{madelon.hulsebos@cwi.nl}}}
\\
 \textsuperscript{1}University of Amsterdam, \textsuperscript{2}CWI
}

\begin{document}

\maketitle

\begin{abstract}
The emergence of open data portals necessitates more attention to protecting sensitive data before datasets get published and exchanged.
To do so effectively, we observe the need to refine and broaden our definitions of sensitive data, and argue that \textit{the sensitivity of data depends on its context}. Following this definition, we introduce a contextual data sensitivity framework building on two core concepts: 1) \textit{type contextualization}, which considers the type of the data values at hand within the overall context of the dataset or document to assess their true sensitivity, and 2) \textit{domain contextualization}, which assesses the sensitivity of data values informed by domain-specific information external to the dataset, such as geographic origin of a dataset. Experiments instrumented with language models confirm that: 1) type-contextualization significantly reduces the number of false positives for type-based sensitive data detection and reaches a recall of 94\% compared to 63\% with commercial tools, and 2) domain-contextualization leveraging sensitivity rule retrieval effectively grounds sensitive data detection in relevant context in non-standard data domains. A case study with humanitarian data experts also illustrates that context-grounded explanations provide useful guidance in manual data auditing processes. We open-source the implementation of the mechanisms and annotated datasets at \url{https://github.com/trl-lab/sensitive-data-detection}.

\end{abstract}

%
%
%
\section{Introduction}  


Data sharing underpins data reuse across sciences, enterprises, and governments through public and private portals~\citep{borgman2025data, worth2024datatransparency, zhang2024datamarketssurvey, wilkinson2016fair, brickley2019googledataset}. With the rise of open data sharing portals, comes the need to protect sensitive data from undesired and harmful use.
Sensitive data includes direct personal identifiers (e.g., names, emails~\citep{gdpr}) as well as indirect or contextual information that becomes a potential risk under certain conditions~\citep{unocha2025}. For example, facility locations or military logs may be sensitive depending on geopolitics or timing. This illustrates the complexity of sensitive data detection and the need for nuanced, context-dependent methods.
Beyond operational and ethical concerns of publishing sensitive data, Large Language Models (LLMs) increase the risks of leakage: trained on web data, they may memorize and reproduce sensitive information~\citep{subramani2023piihuggingface,worth2024datatransparency, carlini2021extracting, lukas2023analyzing}. Studies show sensitive data is often underspecified in documentation on open data portals like HuggingFace~\citep{yang2024navigatingdatasetdocumentationsai,croissant2024benjelloun,worth2024datatransparency}, illustrating the importance of accurate methods and tools for (semi-)automated sensitive data protection, particularly on open portals.

Most work on sensitive data detection methods, however, focuses on detecting person-level data, particularly personal identifiable information (PII)~\citep{KUZINA2023119924,subramani2023piihuggingface,croissant2024benjelloun}. For unstructured data, PII detection relies on named entity recognition~\citep{subramani2023piihuggingface}; similar approaches exist for tabular data~\citep{raman2001potter, Hulsebos2019Sherlock}, which is the focus in this paper. Yet \textit{sensitivity} extends far beyond explicit PII. \citet{kober2023whatissensitive} propose a taxonomy of always-, combination-, context-, and value-sensitive data. For instance, an \texttt{address} may be harmless in the context of a public organizational website but expose a private address in another document, causing na\"{i}ve type-based detection yield false positives. Likewise, hospital \texttt{geo-coordinates} may be harmless in one region but sensitive in conflict zones~\citep{unocha2025}, calling for contextualization beyond detecting personal-sensitive data. These limitations call for a more nuanced perspective and framework for assessing sensitive data.

In this paper, we revisit the concept of sensitive data detection, arguing for a more contextual perspective. We particularly focus on two aspects of contextual sensitive data detection. First, \textit{type contextualization} which takes a more nuanced view beyond the type of values (e.g. \texttt{addresses}) by taking the broader context of the data into account. We introduce the technical ``detect-then-reflect'' mechanism which first detects the type of the values at hand, then reflects on their actual sensitivity based on contextual signals within the document or dataset. Experiments with real-world datasets containing personal identifiable information (PII) show that this mechanism, instrumented by language models, yields a strong improvement in precision while maintaining high recall. Second, we posit \textit{domain contextualization} which addresses sensitivity that depends on external domain-specific factors, beyond merely static types such as PII. We introduce the ``retrieve-then-detect'' mechanism which first retrieves relevant domain-specific information, such as applicable data policies, to then ground the data sensitivity detection in using a language model. Evaluations illustrate the potential of this mechanism, achieving high recall and improved consistency for detecting non-personal sensitive data. Importantly, the context-grounded domain-specific explanations are deemed highly useful by experts in humanitarian data assessment in a case-study.

\section{Revisiting Sensitive Data Detection}\label{sec:revisiting-sensitive-data}


The notion of ``sensitive data'' is multifaceted, spanning legal, theoretical, and domain-specific views. Legal frameworks often equate sensitivity with personal identifiable information. The GDPR defines personal data broadly as ``any information relating to an identified or identifiable natural person,'' specifying sensitive categories like health, ethnicity, or politics requiring extra protection~\citep{gdpr}. Similarly, HIPAA defines identifiers that make health information ``Protected Health Information (PHI)'' when linked to individuals~\citep{hipaa}.
Yet, sensitivity is often contextual: seemingly harmless data can become sensitive when combined, repurposed, or shared across contexts~\citep{malkin2022contextualintegrity}.
Theoretical perspectives deepen this nuance. Nissenbaum’s Contextual Integrity frames sensitivity as the appropriateness of information flows within social norms~\citep{nissenbaum2004privacy}. 
Building on this, researchers advocate context-grounded definitions~\citep{malkin2022contextualintegrity}.
\citet{kober2023whatissensitive} distinguish: (1) always-sensitive data (e.g., personal identifiers); (2) sensitivity by linking data; (3) sensitivity by use; and (4) sensitivity from specific values. This echos that not all personal is equally sensitive, and not all sensitive data is on individuals. To detect sensitive data, of any kind, our framework builds on this conceptual notion of data sensitivity (Def. \ref{def:contextual-sensitive-data}): data protection methods must ensure coverage beyond personal- and static sensitivity definitions.

\vspace{0.15cm}

\begin{definition}[Contextual Sensitive Data]{Data whose sensitivity depends on external factors and requires protection due to how, by whom, and in what context it can be misused.}\label{def:contextual-sensitive-data}
\end{definition}

\vspace{0.15cm}

\noindent Informed by this \textit{contextual} definition of sensitive data, we formalize the problem of contextual sensitive data detection as follows:


\begin{problem}[Contextual Sensitive Data Detection]Given an unordered or ordered sample of values $V = {v_1,v_2,…,v_N}$ from document D, where each value $v_i$ represents a value, the set of internal context factors $C_I$ within $\mathcal{D}$ (e.g., surrounding values, structure), and external context factors $C_E$ outside $\mathcal{D}$ (e.g., specified rules, time/location), the objective of contextual sensitive data detection is to estimate the probability $P(V=sensitive | C_E, C_I) \in [0,1]$, and, based on this probability, classify the sensitivity of V into one of the predefined sensitivity levels ${s_1,s_2,…,s_M}$, where each $s_i$ corresponds to a distinct sensitivity level (e.g., "sensitive", “medium sensitive”, "non-sensitive", etc.).
\end{problem}\label{def:contextual-sensitive-data-detection}

\section{Contextual Sensitive Data Detection}\label{sec:contextual-sensitive-data-detection}

Protecting sensitive data generally follows a three-step procedure: \ding{192} specifying sensitive data, \ding{193} detecting sensitive data, \ding{194} remedying data sensitivity. In this work, we focus on the second step, which has received little attention in research. Existing methods for detecting sensitive data suffer from two key limitations to address the contextual nature of sensitive data:

\begin{enumerate}[topsep=0pt,itemsep=0ex,partopsep=1ex,parsep=1ex]
    \item \textit{Insufficient specificity}: sensitive data detection methods treat all values of types of interest as equally sensitive, while the subject of the value implies its sensitivity. This indistinctness leads to low precision hence high false positive rates.
    \item \textit{Insufficient coverage}: sensitive data detection methods typically focus on personal-level data, particularly PII, and do not generalize to domain-specific sensitive data. The lack of coverage leads to low recall of methods, hence leaving sensitive data unprotected.
\end{enumerate}


To address these challenges in contextual sensitive data detection, we posit two contextual aspects that matter in order to assess the contextual sensitivity of data: \textit{type contextualization} and \textit{domain contextualization}. We introduce two mechanisms to consider these contextual aspects of sensitive data: the \textit{detect-then-reflect} mechanism for type contextualization and the \textit{retrieve-then-detect} mechanism for domain contextualization. In what follows, we specifically focus on tabular data but these mechanisms extend to other modalities, such as free-form text documents.

\subsection{Type Contextualization}

Following existing standards and protocols~\cite{gdpr, hipaa}, sensitive data detection methods often focus on detecting data \textit{types} reflecting personal sensitive data (e.g., names, addresses)~\cite{dlp}. But the appearance of a name or address does not immediately make a data value sensitive. For example, an address may reflect the address of a person or organization, each having significantly different consequences, hence different sensitivity levels. We term this \textit{type contextualization}: whether a data value corresponding to a sensitivity type of interest (e.g. \texttt{name}) truly is sensitive requires considering the context within the document or dataset in which it appears. Type contextualization improves specificity as potential sensitive values are only considered sensitive based on the broader context of the value.


\paragraph{The \textit{detect-then-reflect} mechanism}
Existing methods classify sensitive data based on signals indicating a type of interest, e.g. \texttt{names}, \texttt{addresses}, or \texttt{phone numbers}. If a value matches this type, it is flagged as sensitive. Type contextualization employs a more nuanced assessment, for which we introduce the two-stage detect-then-reflect mechanism for detecting type-informed sensitive data. Algorithm \ref{alg:type-detect-then-reflect} illustrates this staged procedure for binary sensitivity classification.

\begin{algorithm}[h]
\caption{Type Contextualization}
\label{alg:type-detect-then-reflect}
\begin{algorithmic}[1]

\Require Values $V_i=\{v_1,\dots,v_N\}$ from document $D$
\Require Internal context $C_I$, sensitive types $T_{\text{sens}}$, threshold $t$

\State $t_i \gets \textsc{DetectType}(V_i)$

\If{$t_i \in T_{\text{sens}}$}
    \State $p_{\text{sens}} \gets \textsc{Estimate}$ $ P(V_i=\text{sensitive} \mid t_i, C_I)$
    \State $s_i \gets (p_{\text{sens}} > t)$
\Else
    \State $s_i \gets$ \text{false}
\EndIf
\end{algorithmic}
\end{algorithm}
\vspace{-0.2cm}

\noindent In the initial \underline{detection} step, the potential sensitivity of a set of values is estimated by first classifying the values ($V_i$) corresponding to sensitive types of interest  specified upfront ($T_\text{sens}$). This step prioritizes recall by capturing all values that may carry data of the sensitive type. The purpose of the \underline{reflection} step is to estimate the actual sensitivity of the values ($p_\text{sens}$) given their detected sensitive type ($t_i$) and internal context of the full document ({$C_I$}), to finally turn into a discrete sensitivity label ($s_i$). This step is crucial for mitigating false positives from the type detection stage by incorporating inter-value relationships and document semantics, to filter out values of sensitive types that are not actually sensitive.

\begin{figure}[h]
    \centering
    \includegraphics[width=0.88\linewidth]{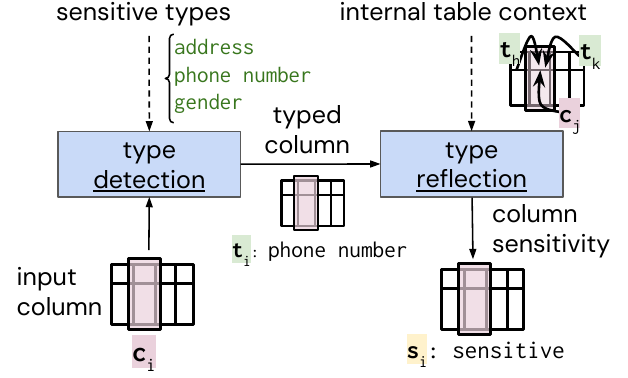}
    \vspace{-0.15cm}
    \caption{\textit{Type contextualization} for contextually detecting sensitive table columns using the \textit{detect-then-reflect} mechanism to increase precision. This mechanism entails: \ding{192} \textit{detecting} potentially sensitive columns based on their (semantic) type, \ding{193} \textit{reflecting} on the actual sensitivity of the column within the internal context of the entire table.}
    \label{fig:type-detect-then-reflect}
    \vspace{-0.25cm}
\end{figure}

\paragraph{Tabular Data case.} Figure~\ref{fig:type-detect-then-reflect} illustrates the detect-then-reflect mechanism for type contextualization on a concrete example with tabular data. Here, table columns are classified to correspond to any sensitive types such as \texttt{address} and \texttt{phone number} based on the column name and data values. For columns of sensitive types, the \textit{reflection} step assesses whether the column is sensitive in context of the full table context, i.e., the table name as well as column names and values of surrounding columns. 

\subsection{Domain Contextualization}
Besides a more fine-grained contextual notion of the sensitivity of certain values, we also argue for a more coarse-grained contextual perspective. Following our notion of contextual sensitive data (Def. \ref{def:contextual-sensitive-data}), we pose that data protection should account for risks beyond identity exposure, for example, harming vulnerable populations or organizations hence consider the data domain, from humanitarian aid to national security. We refer to this broader perspective on data sensitivity as \textit{domain contextualization}: sensitivity emerges from the dataset’s nature and potential misuse, which can arise from external factors. As such, domain contextualization aims at higher detection coverage by grounding the sensitivity assessment in domain-specific context, beyond generic and static predefined types.


\paragraph{The \textit{retrieve-then-detect} mechanism}
While type-based data sensitivity can often be inferred from internal document context, domain-specific sensitivity typically depends on contextual factors external to the data itself. For domain contextualization, we introduce the \textit{retrieve-then-detect} mechanism (Alg. \ref{alg:retrieve-then-detect}).

\begin{algorithm}[h]
\caption{Domain Contextualization}
\label{alg:retrieve-then-detect}
\begin{algorithmic}[1]
\Require Values $V_i=\{v_1,\dots,v_N\}$ from document $D$
\Require External context $C_E$, threshold $t$

\State $C_i^{\text{relevant}} \gets \textsc{RetrieveContext}(V_i, C_E)$
\State $p_{\text{sens}} \gets \textsc{Estimate}$ $P(V_i=\text{sensitive} \mid C_i^{\text{relevant}})$
\State $s_i \gets (p_{\text{sens}} > t)$

\end{algorithmic}
\end{algorithm}
\vspace{-0.15cm}

\noindent Unlike type-informed detection, based on a predefined set of sensitive types, domain-specific sensitive data has no fixed ontology. Instead, the retrieve-then-detect mechanism considers external contextual information, such as rule-based heuristics specified in data governance documents or geopolitical news items, in the detection process. In the \underline{retrieval} stage, external context is selected from a corpus of domain-specific documents or other external data sources ($C_E$) based on the relevance to the potential sensitive values at hand ($V_i$). In the \underline{detection} stage, the sensitivity of the values at hand ($s_i$) is assessed given the retrieved relevant external context ($C_i^\text{relevant}$).

\begin{figure}[b]
    \centering
    \vspace{-0.1cm}
    \includegraphics[width=\linewidth]{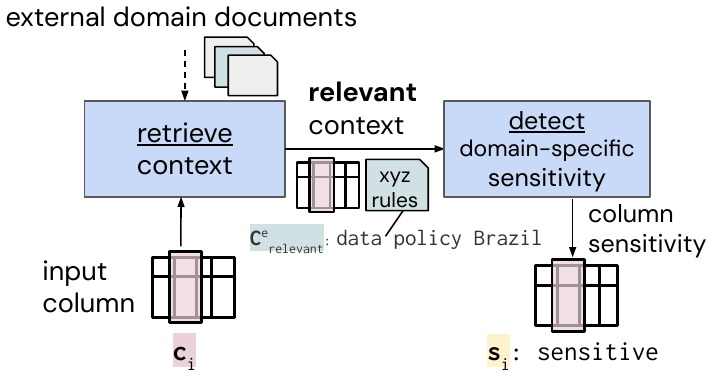}
    \caption{The \textit{retrieve-then-detect} mechanism for detecting sensitive data that requires domain contextualization. This mechanism entails: \ding{192} retrieving relevant contextual information such as data policy rules, \ding{193} detecting sensitive data based on the table and the retrieved relevant context.}
    \label{fig:non-personal-retrieve-then-detect}
\end{figure}

\paragraph{Tabular Data case.} Figure~\ref{fig:non-personal-retrieve-then-detect} illustrates the retrieve-then-detect mechanism for tabular data. For a given column (or table), the retrieval component fetches information, such as data policy rules or geopolitical information, from external documents or other data sources that are relevant for the given input table. The detection step then assesses whether a table (column), or combination of columns, contains data that should be considered sensitive given the retrieved relevant context.

\section{Experimental Setup}
Here, we outline the method implementation details, baselines, datasets, and metrics to evaluate our proposed framework for contextually sensitive data detection. Code and datasets can be found at \url{https://github.com/trl-lab/sensitive-data-detection}.

\subsection{Implementation of the Mechanisms}

The core components of the mechanisms are type detection, reflection, and retrieval. Further details of the implementation of these components per mechanism are provided below.

\paragraph{Type-contextualization}

In the \underline{detection} step, given the well-defined nature of PII types, we parse the problem as a multiclass classification task over a taxonomy of personal identifiable types (e.g., \texttt{email\_address}, \texttt{generic\_identifier}) derived from GDPR \cite{gdpr} and UN OCHA guidelines \cite{OCHA2021Guidelines} (see the code repository for a complete list). We devise a language model to classify the type of a table column (or return \texttt{None})~\cite{kayali2024chorus,shen2025enhancingdeidentificationpersonallyidentifiable}, by prompting it with the column name and five values. We consider zero-shot and fine-tuned approaches. The zero-shot method leverages the model's inherent ability to follow instructions, while the fine-tuned approach enhances performance by training the models on a synthetic dataset of 1,000 labeled columns. Fine-tuning is done using Low-Rank Adaptation \cite{LORA21_HU} for parameter efficiency. The models are trained for a single epoch using an 80/20 train-validation split. 

In the \underline{reflection} phase, the LLM re-evaluates each candidate column corresponding to types of interest within the context of the entire table. We provide the LLM with a markdown-formatted representation of the full table, including headers and five sample rows per column, along with the specific type detected for the target column and, if applicable, the types of other columns which are detected as PII types. The LLM is then instructed to assign one of three contextual sensitivity levels: \texttt{non\_sensitive}, referring to information that cannot identify a person (e.g., aggregate data, organization names); \texttt{moderate\_sensitive}, referring to data that could potentially identify a person when combined with other attributes (e.g., demographics, partial information); and \texttt{high\_sensitive}, referring to data that definitively identifies a person (e.g., full name, email, national ID).

\paragraph{Domain-contextualization}
Given the domain-specific nature of this mechanism, we conduct a case study with humanitarian data experts from the UN OCHA Humanitarian Data Centre using humanitarian crisis datasets from the Humanitarian Data Exchange (HDX) \cite{hdx}. For the \underline{retrieval} step, we aim to identify relevant heuristics indicating sensitivity rules from domain-specific data governance documents. The system retrieves relevant policy fragments from Information Sharing Protocols (ISPs) based on the dataset's geographical context (e.g., country). We use GPT-4o-mini to extract and index the specific sensitivity rules from these documents. If no region-specific ISP is identified, a default ISP template based on cross-country consensus is used.

In the \underline{detection} phase, the LLM is provided with a curated prompt containing table context (headers and five representative rows), relevant ISP extracts containing explicit instructions to classify each column's sensitivity according to the ISP guidelines while justifying its decision by directly citing the relevant policy passages. Based on the policy definitions, the model assigns one of four sensitivity levels, \texttt{non\_sensitive}, \texttt{moderate\_sensitive}, \texttt{high\_sensitive}, or \texttt{severe\_sensitive}. This mechanism ensures that sensitivity assessments are grounded in explicit domain policies, thereby enhancing both interpretability and practical usability for domain experts.


\subsection{Model Selection}
\label{sec:model-selection}
We evaluate closed-source and open-source models to understand trade-offs between these options, if any. While one of the goals of this work is to prevent sensitive data from leaking into proprietary LLMs, these LLMs are typically available under privacy-preserving protocols at some cost which can then provide a viable low-maintenance solution. From closed-source options, we select the GPT-4o-mini~\citep{openai2024gpt4ocard} model for its good performance and cost trade-off. We focus on open-source LLMs that enable local deployment, for which we select the Gemma family (2 and 3)~\citep{gemma_2025,Gemma2Paper} and Qwen3 in 8B and 14B sizes~\citep{yang2025qwen3technicalreport}. Qwen3 14B was selected given the strong performance of its predecessor Qwen2.5's exceptional performance on benchmarks such as MMLU Pro~\footnote{\url{https://huggingface.co/spaces/TIGER-Lab/MMLU-Pro}}, demonstrating strong reasoning capabilities and robust general understanding. Furthermore, Gemma2 9B and Gemma3 12B were included to strike a balance between computational efficiency and performance, and their support for quantized deployment.
We also include versions of Qwen3 8B and Gemma2 9B that are fine-tuned on a synthetic dataset of 1K table columns hand-labeled with 27 PII types, indicated with ``FT'' suffix. 

\subsection{Baselines}
We compare our methods with two common baselines for sensitive data detection. We include Google Cloud DLP \cite{google_DLP}, a rule-based system that combines built-in \textit{infoTypes}, predefined detectors for common Personal Identifiable Information (PII) types, which are associated custom-defined dictionaries and regular expressions. For our experiments, we configure the minimum likelihood threshold to ``Likely''. The second baseline is Microsoft Presidio, an open-source tool that leverages regular expressions, pattern-based rules, context windows, and ML-based named entity recognition to detect PII types. To enable consistent evaluation, each Presidio type has been mapped to a corresponding category in our unified taxonomy (this mapping can be found in the code repository). Both systems output column-level predictions and are widely adopted in enterprise settings. However, neither tool supports contextual reflection or table-wide reasoning; any match to a known PII pattern is automatically treated as sensitive.

\subsection{Datasets}
To evaluate the contextual sensitive data detection mechanisms we use a combination of synthetic and real-world datasets, to maintain ground-truth validity for sensitive data detection in generic context and the domain-specific context of humanitarian datasets.

\begin{itemize}[leftmargin=1em]
    \item \textbf{Type contextualization with sensitive PII}: We extracted 66 anonymized tables, with in total 2,061 columns, from the GitTables corpus \cite{hulsebos2023gittables} representing a variety of real-world table schemas. Each column was manually annotated with (i) its PII type and (ii) its contextual sensitivity level (non-sensitive, moderate, or highly sensitive).

    \item \textbf{Domain contextualization with humanitarian data}: To evaluate the retrieve-then-detect mechanism for domain-specific sensitive data, we conduct a case study with humanitarian experts from the UN OCHA Humanitarian Data Centre. We collect 23 tabular datasets, with in total 326 columns, resembling typical humanitarian datasets (e.g., for displacement tracking, public health monitoring) for different regions. With rich input from humanitarian experts to ensure realistic properties, we synthesize 9 tabular datasets with one or more actual contextually sensitive data in line with region-specific Information Sharing Protocols (ISPs). Two humanitarian data experts annotated columns (100 in total, of which 70 contain sensitive data) with a sensitivity level (low, moderate, high, severe). The other 14 tabular datasets are similar but real tabular datasets from the UN's Humanitarian Data Exchange\footnote{\url{https://data.humdata.org/}} and have been manually curated as challenging datasets without actual sensitive data, to evaluate the false positive rate.
\end{itemize}

\paragraph{Evaluation granularity} For evaluation, sensitivity levels are binarized: \emph{low} is treated as non-sensitive, while \emph{moderate}, \emph{high}, and \emph{severe} are grouped as sensitive. This aligns with operational practice in humanitarian contexts where any moderate or higher sensitivity triggers safeguards, consistent with UN OCHA guidance that such data warrants protective measures~\citep{unocha2025}. For domain contextualization, we assess performance on table-level to align with the ISP's coarse-grained data policy instructions, despite having column-level annotations. Tables are labeled sensitive if one of its columns is annotated as sensitive (moderate, high or severe).

\subsection{Metrics}

We evaluate accuracy along three axes: PII detection, contextual reflection for PII, and non-PII sensitivity detection. To account for class imbalance, we report (support-)weighted averages. Performance is measured at column and table level. For type contextualization we evaluate column-level accuracy, PII detection is treated as a multiclass task over 27 PII types (plus a ``None'' category), while sensitivity reflection reduces to a binary decision between sensitive (moderate/high) and non-sensitive (low). For domain contextualization we evaluate table-level accuracy: a table is classified as sensitive if it contains any moderate/high/severe sensitive information, resulting in a binary task (sensitive vs. not). Precision, recall, and F1 are used as evaluation metrics.

Besides accuracy metrics, we report the average runtime per column to assess deployment considerations.

\section{Results}

We present a comprehensive evaluation of our proposed framework for contextually sensitive data detection in tabular datasets, through type- and domain-contextualization. Type Contextualization experiments instrumented by the detect-then-reflect mechanism are focused on personal-identifiable information (PII) and Domain Contextualization instrumented by the retrieve-then-detect mechanism which focuses on domain-specific sensitivity through a case-study with humanitarian data experts.

\subsection{Type contextualization improves specificity.}

First, we assess the \textit{detect-then-reflect} mechanism for detecting sensitive data through type contextualization. We report on baselines representing existing state-of-the-art tools for detecting sensitive data which automatically treat all columns of interest types as sensitive above the line. Below the line, we report the contextual LLM-assisted method that detects columns of the types of interest which are directly labeled as sensitive \textit{without reflection} (left column), and are labeled as sensitive only if this is the output of the \textit{reflection} step (right column).

\begin{table}[h]
    \centering
    \caption{Type contextualization for PII sensitive data detection, with baselines and the \textit{detect-then-reflect} mechanism. The LLMs are shown with and without the reflection step. We observe that reflection consistently improves precision and F1.}
    \label{tab:sensitivity-evaluation-pii-withwithout-reflection}
    \renewcommand{\arraystretch}{1.2}
    \resizebox{\columnwidth}{!}{%
    \begin{tabular}{lccc|ccc}
    \toprule
         \multirow{2}{*}{\textbf{system/model}} 
         & \multicolumn{3}{c|}{\textbf{w/o reflection}} 
         & \multicolumn{3}{c}{\textbf{w/ reflection}} \\
         & \textbf{prec.} & \textbf{rec.} & \textbf{F1} 
         & \textbf{prec.} & \textbf{rec.} & \textbf{F1} \\
    \midrule
        \textsc{Google DLP} & 0.53 & 0.63 & 0.58 & -- & -- & -- \\
        \textsc{Presidio}   & 0.52 & 0.62 & 0.57 & -- & -- & -- \\
    \midrule
        \textsc{GPT-4o-mini} & \textbf{0.86} & 0.64 & 0.73 & \textbf{0.94} & 0.63 & 0.76 \\
        \textsc{Gemma 2 9B} & \underline{0.74} & 0.82 & \underline{0.78} & 0.80 & 0.79 & 0.80 \\
        \textsc{Gemma 3 12B} & 0.49 & \underline{0.94} & 0.64 & 0.75 & 0.81 & 0.78 \\
        \textsc{Qwen3 8B}   & \underline{0.74} & 0.87 & \textbf{0.80} & 0.75 & \underline{0.87} & 0.80 \\
        \textsc{Qwen3 14B}  & 0.57 & \textbf{0.97} & 0.71 & 0.73 & \textbf{0.94} & \underline{0.82} \\
        \footnotesize{\textsc{Qwen3 8B} FT -> \textsc{GPT-4o-mini}} & -- & -- & -- & \underline{0.90} & 0.86 & \textbf{0.88} \\
    \bottomrule
    \end{tabular}
    }
\end{table}

Without reflection, LLMs mark any detected PII-type column as sensitive, leading to high recall but lower precision; Gemma 3 12B and Qwen3 14B reach a recall of 0.94 and 0.97 but precision of only 0.49 and 0.57 respectively (Table ~\ref{tab:sensitivity-evaluation-pii-withwithout-reflection}). GPT-4o-mini achieves higher precision, 0.86, at lower recall of 0.64. With reflection using full-table context, substantially improves precision across models, e.g., Qwen3 14B increases to 0.73 precision with 0.94 recall, trading a small decrease in recall for a significant gain on precision. A modular pipeline combining Qwen3 8B for type detection with GPT-4o-mini for reflection achieves F1 0.88, but is closely matched by Qwen3 14B which yields higher recall (0.94 compared to 0.86). These results illustrate the stark difference compared to existing baselines reported in the upper left quadrant, which suffer from insufficient recall \'{a}nd precision. Comparing the detect-then-reflect mechanism powered by open-source LLMs with commonly used pattern-matching tools DLP and Presidio, we observe a significant improvement in both precision and recall, addressing the key limitations of existing tools for deployment in practice. Overall, the \textit{detect-then-reflect} mechanism corrects the naive assumption that all columns of sensitive \textit{types} are indeed always sensitive, boosting precision while maintaining recall.

\paragraph{Ablation: reflection without detection.}

To investigate the value of the type-detection step, we also task models with identifying sensitive columns using directly the full table context without first detecting the presence of columns of specific types (Table \ref{tab:sensitivity-evaluation-pii-withwithout-reflection}). The results in Table~\ref{tab:PII-no-detection-reflection} illustrate that LLMs yield good performance for sensitive columns using only table context without prior type detection, but generally yield significant improvement across models when explicit types are detected first. Qwen3 14B achieves the highest F1 (0.82) followed by Gemma 2 9B (F1 0.81). GPT-4o-mini excels in precision (0.96) but has low coverage with a recall of 0.56. Comparing the performance with the detect-then-reflect pipeline in Table \ref{tab:sensitivity-evaluation-pii-withwithout-reflection}, we observe that the reflection-only approach, without type detection, yields particularly a lower coverage, therefore leaving more sensitive data undetected. For example, looking closer into the performance of the best performing models, the recall of Qwen3 14B drops from 0.94 to 0.89, and Gemma2 9B from 0.79 to 0.75. These results indicate that LLMs can reason about sensitivity from the table alone, but the full \textit{detect-then-reflect} mechanism yields preferable trade-off in precision and recall.

\begin{table}[h!]
    \centering
    \caption{Reflection-only performance without prior type detection, for identifying contextually sensitive columns. Evaluated across models using precision, recall, and F1 scores.}
    \begin{tabular}{lccc}
    \toprule
         \textbf{model} &  \textbf{precision} &  \textbf{recall} &  \textbf{F1}\\
         \toprule
         \textsc{GPT-4o-mini} &\textbf{ 0.96}   & 0.56 & 0.71  \\
         \textsc{Gemma 2 9B} & \underline{0.88}   & 0.75 & \underline{0.81}   \\
         \textsc{Qwen3 8B} & 0.74   & \underline{0.78}& 0.76  \\
         \textsc{Qwen3 14B} & 0.76   & \textbf{0.89}& \textbf{0.82} \\
     \bottomrule
    \end{tabular}
    \label{tab:PII-no-detection-reflection}
    \vspace{-0.2cm}
\end{table}

\subsection{Domain contextualization provides coverage and context-grounded explainability.}

Shifting towards contextual sensitive data which cannot be assessed through a fixed set of types, we evaluate the \textit{retrieve-then-detect} mechanism on a combined set of synthetic and real humanitarian tables from the HDX platform. Without external domain knowledge, all models exhibit a highly conservative bias, achieving perfect recall but all with precision lower than 0.56 (Table \ref{tab:non-PII-results}; baseline). Integrating retrieved domain-specific context, i.e., sensitivity guidelines obtained from information sharing protocols (ISP's) for humanitarian data sharing, into the input for reflection by the language model, consistently improves precision. GPT-4o-mini's precision increases from 0.47 to 0.69 (F1=0.82), and Qwen3 14B's from 0.56 to 0.64 (F1=0.78), both maintaining perfect recall (1). Table \ref{tab:non-PII-results} summarizes the results on the 23 datasets humanitarian datasets.

\begin{table}[h]
    \centering
    \caption{Table-level classification performance with and without context-grounding in humanitarian data sensitivity rules. The retrieved rules improve precision while maintaining high recall.}
    \label{tab:non-PII-results}
    \renewcommand{\arraystretch}{1.2}
    \resizebox{\columnwidth}{!}{%
    \begin{tabular}{lccc|ccc}
    \toprule
         \multirow{2}{*}{\textbf{system/model}} & \multicolumn{3}{c|}{\textbf{w/o domain knowledge}} & \multicolumn{3}{c}{\textbf{w/ domain knowledge}} \\
         & \textbf{prec.} & \textbf{rec.} & \textbf{F1} & \textbf{prec.} & \textbf{rec.} & \textbf{F1} \\
         \midrule
         \footnotesize{Baseline: all sensitive} & 0.375 & \textbf{1.000} & 0.545 & -- & -- & -- \\
         \midrule
         \textsc{GPT-4o-mini} & 0.474 & \textbf{1.000} & 0.643 & \underline{0.692} & \textbf{1.000} & \textbf{0.818} \\
         \textsc{Gemma 2 9B} & 0.375 & \textbf{1.000} & 0.545 & 0.429 & \textbf{1.000} & 0.600 \\
         \textsc{Gemma 3 12B} & \underline{0.529} & \textbf{1.000} & \underline{0.692} & 0.500 & \textbf{1.000} & 0.667 \\
         \textsc{Qwen3 8B} & \textbf{0.562} & \textbf{1.000} & \textbf{0.720} & \textbf{0.778} & \underline{0.778} & 0.778 \\
         \textsc{Qwen3 14B} & \textbf{0.562} & \textbf{1.000} & \textbf{0.720} & 0.643 & \textbf{1.000} & \underline{0.783} \\
    \bottomrule
    \end{tabular}
    }
\end{table}

Qualitative analysis surfaces that the context retrieval mechanism enables models to produce interpretable and consistent justifications, thereby enhancing the transparency and auditability of the sensitivity assessments. The model outputs containing sensitivity classifications and explanations were reviewed by two humanitarian data quality experts, who considered the context-grounded explanations accurate, and highly informative for guiding and standardizing human-assessment. The below excerpt shows such context-grounded LLM assessment on a table titled ``Reported location on NFIs suppliers'' with (1) a description of the data at hand, along with (2) the LLM's sensitivity assessment, and (3) explanation grounded in retrieved ISP rules. 

\vspace{0.25cm}

{\small
\begin{mdframed}[backgroundcolor=teal!10,rightline=true,leftline=true,font=\ttfamily, innerrightmargin=0.1cm, innerleftmargin=0.3cm, innertopmargin=0.2cm, innerbottommargin=0.2cm]

1. The column ``nfi\_supplier\_loc-value-from\_Iran'' indicates the
origin of NFIs suppliers.

\vspace{0.15cm}

\noindent 2. This information, even with country-level
disaggregation, doesn't reveal personal or sensitive
data. It relates to the location of suppliers, which is
generally not considered sensitive.

\vspace{0.15cm}

\noindent 3. It aligns with the ISP’s guideline on ``Facility data
(e.g., health, education, water points) at national or
regional level, unless explicitly restricted by clusters.''
Since this column is about locations of suppliers
and not about individuals or specific operations, it
is categorized as non-sensitive.

\end{mdframed}}
\vspace{0.2cm}

\subsection{Error analysis}

\paragraph{Type contextualization} On data with PII, we particularly observed misclassification of columns of generic types, as well as explicit personal types. For example, the ``GENERIC\_ID'' type is well detected, but classified as non-sensitive by most LLMs causing very low recall. For types more explicitly relating to personal information (like names), columns are often classified as sensitive in reflection, while not always annotated as such. This highlights the potential for more specific context- and type-specific instructions during reflection.

\paragraph{Domain contextualization} In our experiments with humanitarian data, we observed mainly false positives for LLMs, which generally appear cautious with humanitarian data. On these false positives, analysis surfaces that the Qwen model has most confusion around borderline cases that are annotated as non-sensitive but classified as medium-sensitive, while gpt-4o-mini deems most of these tables highly sensitive (75\%). This illustrates the need for more specific instructions for borderline cases.

\subsection{Latency of contextual sensitive data detection.}

Table \ref{tab:latency_cost_comparison} gives an overview of wall-clock latency measurements of the end-to-end pipeline averaged over all columns, as well as cost indications and memory measurements for local models. Despite the more elaborate mechanisms for contextualized sensitive data detection, the latency delta between the contextual sensitive LM-powered pipelines versus Google DLP (which only conducts type-based sensitivity detection) might be acceptable given the significant accuracy gain that the introduced mechanisms provide. If efficiency is key, smaller models like Qwen3 8B can yield an favorable trade-off, whereas larger or more costly models for more sensitive scenarios can be feasible for smaller-scale evaluations.

\begin{table}[h]
\centering
\caption{Comparison of pipeline latency per column and resource cost across different methods and models.}
\label{tab:latency_cost_comparison}
\renewcommand{\arraystretch}{1.2}
\resizebox{\columnwidth}{!}{%
    \begin{tabular}{lcc}
    \toprule
    \textbf{tool/model} &  \textbf{total pipeline (s/col)} &  \textbf{cost or GPU memory} \\
    \midrule
    Google DLP\tablefootnote{For Google DLP, we only measure type-based sensitivity detection as it is not suitable for domain-specific sensitivity.}   & 0.16 & subscription \\
    GPT-4o-mini   & 1.20 & \$0.05 per table \\
    Gemma 2 9B    & 2.89 & 7.8 GB \\
    Gemma 3 12B   & 2.94 & 8.0 GB \\
    Qwen3 8B      & 0.46 & 8.0 GB \\
    Qwen3 14B     & 1.62 & 11.4 GB \\
    \bottomrule
    \end{tabular}
}
\end{table}

\section{Related Work}

Existing methods for detecting sensitive data concentrate on detecting PII types, or otherwise static sensitive types. To this end, widely adopted industry tools like Microsoft Presidio \citep{presidio} and Google Cloud DLP \citep{google_DLP} detect correspondences between values and structured types, such as email addresses and phone numbers, based on rule-based heuristics and syntactic pattern matching. Recent research has expanded the type detection scope by utilizing large neural models to learn mappings between high-dimensional semantic representations of values, e.g. table columns, and their semantic types~\cite{Hulsebos2019Sherlock, KUZINA2023119924}.
Language models have also emerged as powerful tools for sensitive data detection, as particularly shown for PII detection, outperforming both rule-based and traditional machine learning methods in accuracy and adaptability~\citep{shen2025enhancingdeidentificationpersonallyidentifiable, Yang2023Exploring}. However, while language models excel at identifying known PII types, they struggle with assessing \textit{contextual} sensitivity, such as domain-specific sensitive types, unless explicitly provided with the relevant operational context.
In this work, we address this gap: our contextual sensitive data detection framework addresses both personal and non-personal sensitive data by considering internal and external context factors. 

\section{Conclusion}

With the rise of open data portals to democratize data access, comes the risk of exposing sensitive data to malicious actors or language models fueled by public data. Protecting sensitive data is therefore key, but methods for this are limited: \textit{they are too specific to capture all sensitive data, and too generic to only suppress data that is actually sensitive}. We argue for \textit{contextualizing} sensitive data detection along two dimensions. First, \textit{type contextualization}, which considers fine-grained aspects beyond static sensitivity types. Our detect-then-reflect mechanism shows that first detecting if data values belong to a sensitive type, then reflecting on their true sensitivity, increases recall over existing tools with up to 49\%. Second, we introduce \textit{domain contextualization} which considers domain-specific factors external to the data in order to assess its sensitivity. A case study with humanitarian datasets highlights that contextualizing the sensitivity assessment in relevant domain-specific rules through a retrieval stage, extends the type coverage and enables domain-grounded explanations. We believe that our work provides a stepping stone in this important direction.

\section*{Limitations and Risks}


\paragraph{Evaluation coverage} In our humanitarian case study, we show that domain contextualization leads to higher precision while preserving coverage, and also provides useful grounded explanations to data quality reviewers. The limited availability of labeled data in this case study, might affect generalization of the insights for domain-contextualized sensitive data detection. Moreover, while the conceptual framework and mechanisms are modality agnostic (Section \ref{sec:contextual-sensitive-data-detection}), the lack of datasets with labeled sensitivity and the data setting of the humanitarian case study have resulted in evaluations for tabular data only, for which we constructed and publish hand-labeled datasets. We call for more diverse labeled datasets, to advance this important problem, and evaluate the framework and mechanisms for different modalities, languages, and domains.

\paragraph{Modality coverage} Despite the focus of our implementation and evaluation on tabular data. The proposed mechanisms easily extend to other modalities, such as free-form text data: 1) type-informed sensitivity in free-form text can be implemented efficiently by concatenating a decomposition component with a more lightweight learned type predictor~\cite{Hulsebos2019Sherlock} and performing reflection in batches, 2) for domain-contextualization of free-form text, a hierarchical reflection procedure could be efficient. For example, starting with assessing high-level segments of the text (document-level, paragraph-level, sentence-level, etc~\cite{lin2025querying}), and incrementally narrowing down.

\paragraph{Sensitivity specification dependency} The presented mechanisms for detecting sensitive data can enable automation but also assist humans in data reviewing by providing context-grounded explanations with explicit citations to relevant sensitivity rules. For successful adoption in practice, however, the mechanisms depend on having access to reasonable sensitivity instructions in the form of 1) fine-grained semantic types of data that are deemed sensitive for type contextualization, and 2) reasonably specified assessment rules in domain-specific documents or (web) data sources.

\paragraph{Data access delay and further automation} A potential negative consequence of automated data sensitivity detection is that critical datasets might be blocked that are time-sensitive and could be instrumental for solving societal problems, for example, in humanitarian domains. The mechanisms could be tailored around the desired precision/recall trade-off to reduce false alarms. In this work, we only present mechanisms for detecting contextual data sensitivity, and not for automated remedies. In practice, a report could be generated based on existing cases to instruct the user on how to remedy the sensitive data, or potentially even automatically remedy this, which is also an important direction for future research.

\section*{Acknowledgements}
We are grateful for the collaboration with the Responsible Data team of the United Nations OCHA Humanitarian Data Centre. Specifically, we thank Alexandru Gartner, Godfrey Takavarasha, Jos Berens, Javier T\'{e}ran, Metasebya Sahlu, Melanie Rabier, Nafissatou Pouye, Sarah Telford, Serban Teodorescu, and Shahrooz Badkoubei, for their invaluable contributions to this project.
This work has been supported by funding from an AiNed grant from NWO (NGF.1607.22.045) and credit grants from OpenAI.

\bibliography{main.bib}

\end{document}